\documentclass[%
 reprint,
 amsmath,amssymb,
 aps,
]{revtex4-2}

\usepackage{graphicx}
\usepackage{dcolumn}
\usepackage{bm}
\usepackage{xcolor}
\usepackage{hyperref}

\usepackage{booktabs}

\begin{document}

\title{Stimulated cooling in non-equilibrium Bose--Einstein condensate}
\author{Ka Kit Kelvin Ho\textsuperscript{1,$\ddagger$}
}

\author{Vladislav~Yu.~Shishkov\textsuperscript{1,$\ddagger$}
}

\author{Mohammad Amini\textsuperscript{1}
}

\author{Leonie Teresa Wrathall\textsuperscript{1}}

\author{Evgeny Mamonov\textsuperscript{2}
}

\author{Darius~Urbonas\textsuperscript{3}}

\author{Ioannis Georgakilas\textsuperscript{4}}


\author{Tobias Herkenrath\textsuperscript{5}}

\author{Michael Forster\textsuperscript{5}}

\author{Ullrich Scherf\textsuperscript{5}}

\author{Tapio Niemi\textsuperscript{6}}

\author{Päivi Törmä\textsuperscript{2}}

\author{Anton~V.~Zasedatelev\textsuperscript{1}}
\email{anton.zasedatelev@aalto.fi}

\affiliation{
\textsuperscript{1}Macroscopic Quantum Optics (MQO) Labs, Department of Applied Physics, Aalto University School of Science, FI-00076 Espoo, Finland
}
\affiliation{\textsuperscript{2}Department of Applied Physics, Aalto University School of Science, FI-00076 Espoo, Finland}
\affiliation{
\textsuperscript{3}IBM Research Europe – Zurich, Säumerstrasse 4, 8803 Rüschlikon, Switzerland}
\affiliation{\textsuperscript{4}Laboratory of Nano and Quantum Technologies, Paul Scherrer Institut, 5323 Villigen, Switzerland}
\affiliation{\textsuperscript{5}Macromolecular Chemistry Group and Wuppertal Center for Smart Materials and Systems (CM@S),
Bergische Universitat, Wuppertal, Gauss Strasse 20, 42119 Wuppertal, Germany}
\affiliation{\textsuperscript{6}Photonics Laboratory, Physics Unit, Tampere University, Korkeakoulunkatu 3, 33720 Tampere, Finland}

\date{January 2026}

\begin{abstract}
We report on the experimental observation of stimulated cooling in the non-equilibrium Bose--Einstein condensate (BEC) of weakly interacting exciton-polaritons from approximately room temperature down to $20~{\rm K}$. By resolving the condensate in energy–momentum space and performing interferometric measurements, we distinguish the condensate from thermalized particles yet occupying excited states macroscopically.
In contrast to the analytical quantum theories of non-equilibrium BEC [Shishkov et al., Phys. Rev. Lett. 128, 065301 (2022)], we observe segmentation of the particle density along the excited states into two fractions both following Bose--Einstein distribution, albeit with different effective temperatures and chemical potentials. Our results indicate that the temperature of the weakly interacting Bose gas is universally set by the density-dependent chemical potential, revealing a defining property of non-equilibrium BECs. Finally, we demonstrate that the stimulated nature of the cooling process directly governs the emergence of quantum coherence of the condensate and shapes the dissipative properties of the excited states.
 
\end{abstract}

\maketitle

\section*{Introduction}
The experimental demonstration of Bose--Einstein condensation (BEC), first achieved in trapped ultracold atoms~\cite{anderson1995observation}, is one of the key milestones of modern physics. 
The BEC is a striking example of collective many-body quantum behavior on a macroscopic scale, sitting alongside earlier landmarks such as superfluidity~\cite{PhysRevLett.28.885, kapitza1938viscosity, allen1938flow} and superconductivity~\cite{onnes1911further, PhysRev.108.1175}.
A light-matter BEC is a collective macroscopic quantum state~\cite{alnatah2024coherence} inheriting strong nonlinearity from the matter component and very small effective mass from optical cavity~\cite{kasprzak2006bose}.
The latter are open quantum systems, with the finite photon lifetime in the cavity, so light–matter BECs are inherently driven–dissipative and, in general, out of thermal equilibrium~\cite{keeling2020bose, bloch2022non}. 
This allows broad range of non-destructive measurement schemes with light–matter BECs, in contrast to ultracold atoms~\cite{bloch2022non}.
Despite the non-equilibrium nature of light-matter condensates, they share many common properties with an ideal Bose gas: particles follow the Bose--Einstein distribution with almost zero chemical potential~\cite{pieczarka2024bose}, spatial coherence emerges spontaneously~\cite{PhysRevLett.115.035301, damm2017first}, and the second-order coherence function approaches unity~\cite{PhysRevB.110.045125}. 
Many of the thermodynamic properties also appear, including cusp singularity of the specific heat~\cite{damm2016calorimetry}, and sharp increase in compressibility at the BEC transition~\cite{busley2022compressibility}. 
The critical particle number follows statistical-physics predictions~\cite{klaers2010thermalization, klaers2010bose}, and the nature of the BEC transition depends on the system dimensionality~\cite{karkihalli2024dimensional}. 
The close analogy with an ideal Bose gas is particularly relevant for weakly-interacting exciton-polations, which can be found in highly bound excitonic systems and are robust against thermal fluctuations at room temperature~\cite{keeling2020bose}. 
The examples are wide band gap semiconductors~\cite{levrat2010condensation, li2013excitonic}, two dimensional materials~\cite{anton2021bosonic, zhao2021ultralow}, molecular systems~\cite{plumhof2014room, hakala2018bose, vakevainen2020sub}, perovskites~\cite{su2020observation}, and others~\cite{schofield2024bose, alnatah2024bose}. 
In fact, for many of these systems BEC has been achieved at room temperature, making them a future building-block for high performance optoelectronic devices~\cite{sanvitto2016road, zasedatelev2019room, zasedatelev2021single, kavokin2022polariton, sannikov2024room}.

Although experiments on non-equilibrium BECs show that many of their properties are closely related to those of an equilibrium ideal Bose gas, we still lack a clear understanding of how these two fundamentally different frameworks, thermodynamics and driven–dissipative dynamics, are connected.
Moreover, the many-body open quantum nature of light-matter BECs grants them a number of unique properties~\cite{bloch2022non}, that are different from the properties of equilibrium BECs~\cite{landau2013statistical}.
Some of these properties, such as spatial distribution of the order parameter~\cite{alyatkin2025quantum, alyatkin2024antiferromagnetic}, build-up of temporal coherence~\cite{kasprzak2006bose, plumhof2014room, alnatah2024bose}, condensation in the excited state~\cite{wertz2010spontaneous, zasedatelev2019room, sannikov2024room}, have been observed, while others are still waiting to be discovered, including stimulated cooling, the anti-correlations between condensed and non-condensed particles at the threshold of BEC, and the rise of the second-order coherence for non-condensed particles at the threshold of BEC~\cite{PhysRevLett.128.065301}.
The lack of understanding of the quantum properties of thermalization in out-of-equilibrium conditions prevents establishing a universal picture behind build-up of coherence and quantum correlations within the process of light-matter condensation.


In this work, we experimentally study thermalization of a weakly interacting Bose gas of exciton-polaritons and observe stimulated cooling from approximately room temperature down to $20~{\rm K}$.
We reveal universal behavior in the dynamics driven by the particle density in the system, which results in two distinct thermalized fractions, having different chemical potentials and temperatures. 
Our results suggest a smooth crossover between these two fractions, connecting non-equilbrium nature of light-matter condensation with the equilibrium physics of the ideal Bose gas.

\section*{The condensate area and excited states}



\begin{figure}
\includegraphics[width=0.89\linewidth]{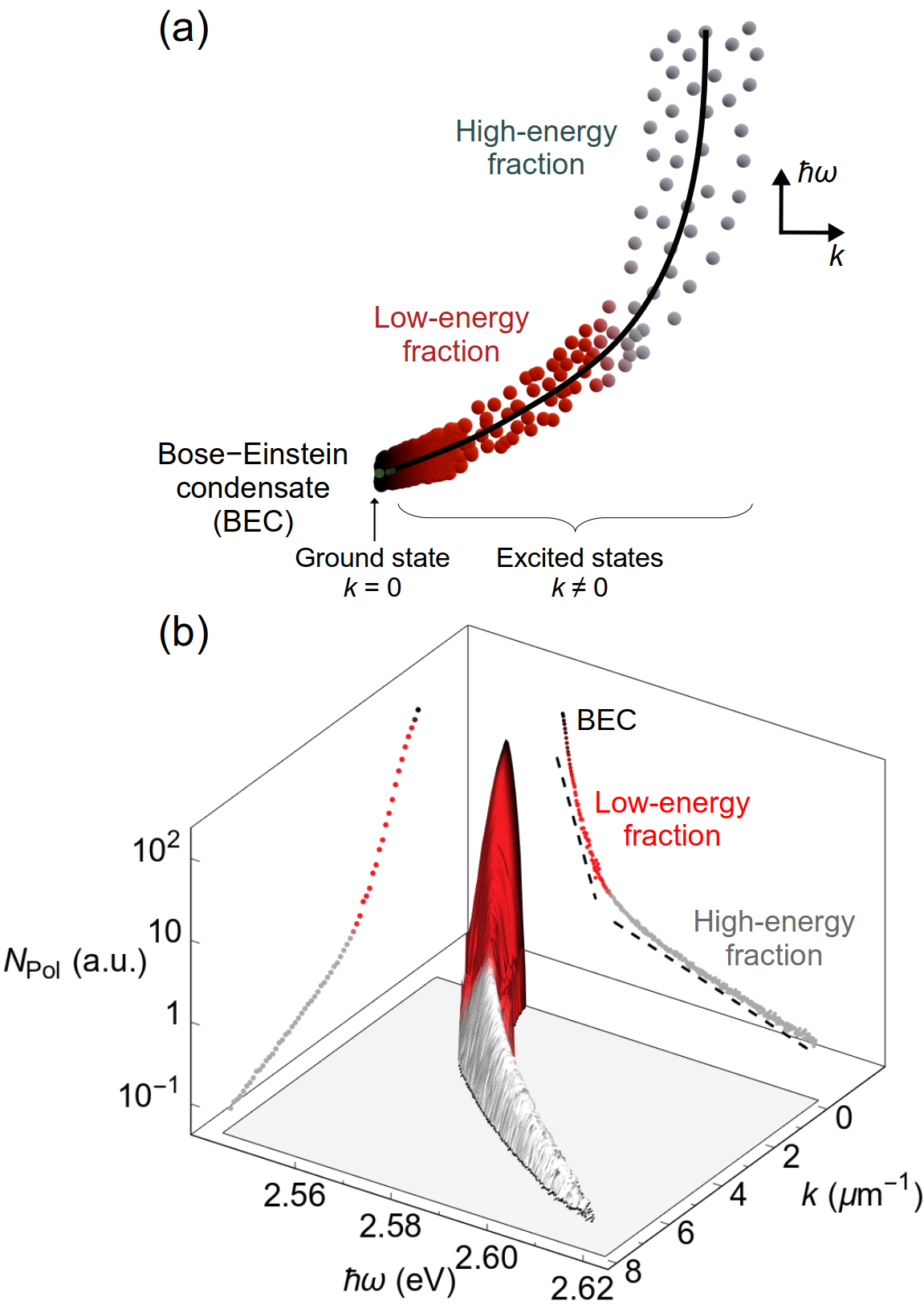}
\caption{
{\bf Nonequilibrium BEC.}
(a) Schematic of the polariton thermalization in the energy-momentum space. 
(b) Particle density distributions as a function of energy and momentum at $P/P_{\mathrm{th}}= 1.6$. 
}
    \label{fig:1}
\end{figure}

We explore thermalization under resonant, blue-detuned excitation conditions~\cite{zasedatelev2021single}. Our material system constitutes of a $\lambda/2$ dielectric Fabry--Perot microcavity with a conjugated polymer as the excitonic medium coupled strongly to the fundamental mode of the cavity. 
This system was among the first to show room-temperature light–matter BEC~\cite{plumhof2014room}. Details on the microcavity structure and the experimental setup are provided in Methods. 
Initially, we populate the lower polariton branch within $k_BT\sim30~{\rm meV}$ range uniformly, following vibrational relaxation from higher-lying polariton states pumped by $50~{\rm fs}$ short pulses~\cite{PhysRevB.110.134321, PhysRevLett.133.186903}.
Thermalization processes rearrange the particles over the dispersion curve, eventually making them follow, first Boltzmann and, then Bose--Einstein distribution~\cite{kavokin2017microcavities}.
The threshold of BEC marks the number of pumped particles at which the Boltzmann distribution changes to the Bose--Einstein distribution.
For two-dimensional Bose gas with quadratic dispersion, this transition from Boltzmann distribution to Bose--Einstein distribution is a crossover rather than abrupt change~\cite{Shishkov2022analyticalframework}. In the ideal Bose gas, the condensate itself represents a quantum coherent macroscopically occupied ground state of the system, while the particles at the higher energies do not exhibit coherent properties and obey thermal distribution (Fig.~\ref{fig:1}a). Figure~\ref{fig:1}b shows a typical \textit{E,k}-distribution of polariton density above condensation threshold. In contrast to the ideal picture, the macroscopic occupation in light-matter BEC appears to be significantly extended towards higher-lying states. At this moment, it is difficult to separate the fraction of the condensate from the thermally distributed particles. Moreover, one can see two distinct fractions in particle distribution across energies, characterized by different effective temperatures (Fig.~\ref{fig:1}b).


\begin{figure}
\includegraphics[width=0.88\linewidth]{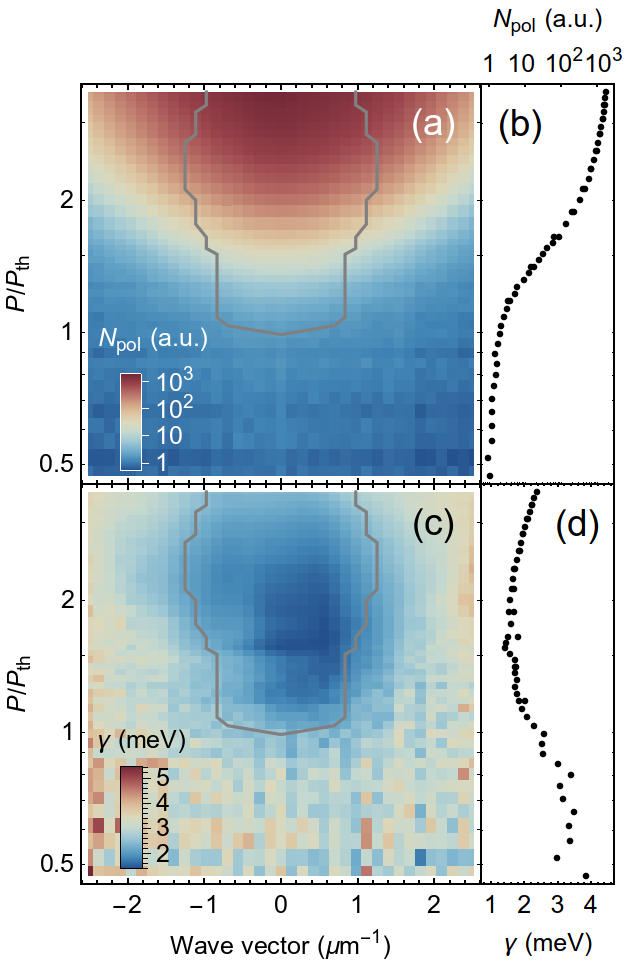}
\caption{
{\bf Condensation area.}
(a) Particle density distribution in momentum space as a function of pump power $P$, normalized to the density at the BEC threshold $P_{\rm th}$.
(b) Ground-state density (${\bf k}={\bf 0}$) as a function of pump power.
(c) Momentum-resolved linewidth as a function of pump power.
(d) Ground-state linewidth as a function of pump power.
The gray line in (a) and (c) indicates the BEC size in momentum space, defined as the momenta ${\bf k}$ such that $\gamma_{\bf k} \leq \gamma_{{\bf k} = {\bf 0}} + (\bar\gamma - \gamma_{{\bf k} = {\bf 0}})/3$, where $\bar\gamma$ is average linewidth below BEC threshold}.
    \label{fig:2}
\end{figure}

The formation of the BEC is accompanied by non-linear increase in the population of the ground state~(Fig.~\ref{fig:2}a,b) and narrowing of the linewidth in the vicinity of the ground state~(Fig.~\ref{fig:2}c,d) corresponding to the build-up of the temporal coherence at these states.
 Above the BEC threshold, the temporal coherence established for the states within entire range of $-1~{\rm \mu m}^{-1}$ to $1~{\rm \mu m}^{-1}$ ~(Fig.~\ref{fig:2}c).
This range in \textit{k}-space remains mostly the same as we pump more particles into the system.
We distinguish the condensate from the thermal gas by defining the condensate area as the region within this range, which sets the BEC size in $k$-space. The coherence build-up within the area results in the emergence of a collective macroscopic quantum state described by a single wave function~\cite{alnatah2024coherence}.
Thus, the finite size of BEC in $k$-space is a defining property of non-equilibrium BEC, contrasting with equilibrium BEC occupying exactly one ground state with zero momentum~\cite{ketterle1996bose}.
Identifying the exact size of the condensate in the momentum/energy space is crucial for the further analysis of quantum dynamics behind non-condensed particles involved in the thermalization physics towards the BEC.



\section*{Segmentation of particles in thermal fractions}

\begin{figure}
\includegraphics[width=1\linewidth]{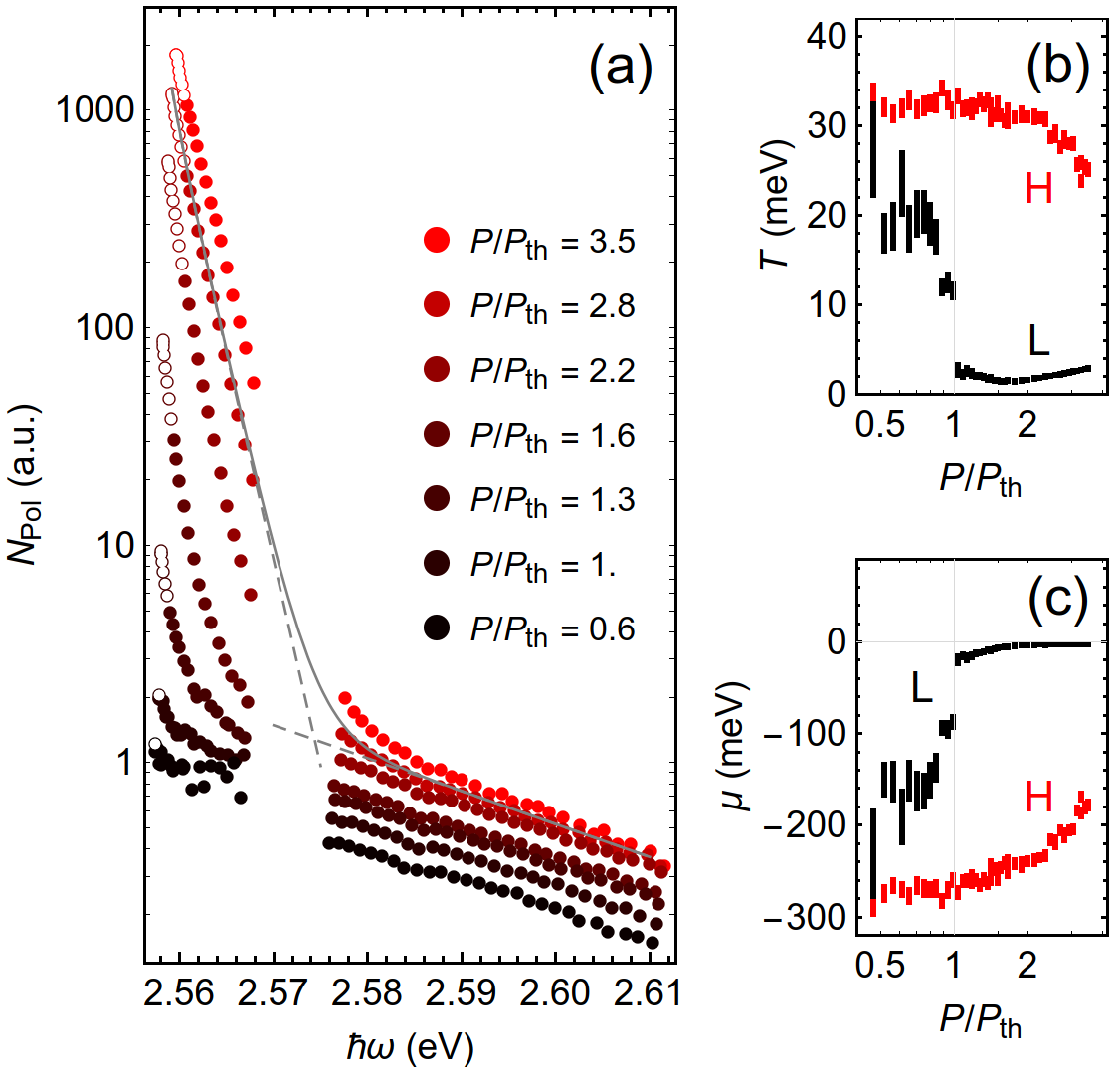}
\caption{
{\bf Spontaneous-to-stimulated thermalization.}
(a) Distribution of particles over energies for different pump powers from below $0.6P_{th}$ to above $3.5P_{th}$ condensation threshold.
Gray solid line is fit with the sum of two Bose--Einstein distributions $N_{\rm Pol} = n^{\rm BE}_{\bf k}(T_{\rm L}, \mu_{\rm L}) + n^{\rm BE}_{\bf k}(T_{\rm H}, \mu_{\rm H})$ at $P/P_{\rm th}=2.8$. 
Gray dashed lines are $n^{\rm BE}_{\bf k}(T_{\rm L}, \mu_{\rm L})$ and $n^{\rm BE}_{\bf k}(T_{\rm H}, \mu_{\rm H})$ at $P/P_{\rm th}=2.8$.
Open dots corresponds the data points which obey condensation area defined in Fig.~\ref{fig:2}. 
(b) Temperature, $T$, and (c) chemical potential, $\mu$, extracted from the fit of the distributions with $n^{\rm BE}_{\bf k}(T_{\rm L}, \mu_{\rm L})$ and $n^{\rm BE}_{\bf k}(T_{\rm H}, \mu_{\rm H})$ for the low-energy range $(2.55~{\rm eV},2.57~{\rm eV})$ (black bars, L) and high-energy range $(2.585~{\rm eV},2.61~{\rm eV})$ (red bars, H) respectively, as functions of pump power normalized to the BEC threshold $P/P_{\rm th}$. 
Error bars indicate the standard error of the fitted temperatures and chemical potentials.
}
    \label{fig:3}
\end{figure}

Above the BEC threshold, the distribution of the particles over the energies becomes steep in a broad range ${\sim 20~{\rm meV}}$ near the ground state~(Fig.~\ref{fig:3}a). 
However, only small fraction of this energy range (${\sim 1~{\rm meV}}$) actually belongs to the condensate area, as indicated by open points in~(Fig.~\ref{fig:3}a).
The remaining particles follow a thermal distribution with an effective temperature as low as $20~{\rm K}$, which accounts for the rapidly decaying tail above the ground state (Fig.~\ref{fig:3}b).
We acknowledge limited range of energies and move into the additional measurements where we blocked the light coming from $\pm 4~{\rm \mu m}^{-1}$, thus physically cutting out the states at the condensate~(Fig.~\ref{fig:3}a). This data clearly reveal the segmentation of the particles around the BEC threshold into two distinct fractions -- low-energy fraction and high-energy fraction -- occupying the states with different energies.
These two fractions follow the Bose--Einstein distribution,

\begin{equation}
n^{\rm BE}_{\bf k}(T, \mu) = \frac{g}{e^{(\hbar\omega_{\bf k}-\hbar\omega_{\bf 0}-\mu)/T}-1},     
\end{equation}
with two different temperatures and chemical potentials, but the same constant $g$, that does not depend on neither $T$ nor $\mu$.
We denote the fitted distributions for the fractions as $n^{\rm BE}_{\bf k}(T_{\rm L}, \mu_{\rm L})$ and $n^{\rm BE}_{\bf k}(T_{\rm H}, \mu_{\rm H})$, where subscripts $\rm L$ and $\rm H$ stand for low-energy fraction and high-energy fraction respectively~(Fig.~\ref{fig:3}b,c).
As we pump more particles into the system, both fractions experience stimulated cooling~(Fig.~\ref{fig:3}b).
Above the condensation threshold the resulting particle distribution can be approximated by $N_{\rm Pol} \approx n^{\rm BE}_{\bf k}(T_{\rm L}, \mu_{\rm L}) + n^{\rm BE}_{\bf k}(T_{\rm H}, \mu_{\rm H})$, which also captures the smooth transition from the low-energy fraction to the high-energy fraction~(Fig.~\ref{fig:3}a).
This separation into two fractions, each with its own temperature and chemical potential, and each undergoing stimulated cooling, is a distinct feature of non-equilibrium BEC. The low-energy fraction is about an order of magnitude colder than the high-energy fraction, yet both exhibit stimulated cooling (Fig.~\ref{fig:3}b), as was predicted within the quantum theory of the nonequilibrium BEC of ideal Bose gas~\cite{PhysRevLett.128.065301}.
As we inject more particles into the system, the chemical potentials of both fractions increase. However, the chemical potential of the low-energy fraction remains substantially higher than that of the high-energy fraction (Fig.~\ref{fig:3}c). This indicates that the total particle density differs across energy ranges, which in turn shapes the thermalization dynamics.
While the chemical potential is monotonous for both fractions~(Fig.~\ref{fig:3}c), the temperature is not, as it starts to grow again for low-energy fraction at high density of particles pumped above $2P_{th}$~(Fig.~\ref{fig:3}b).
We attribute this behavior to the nonlinear effects which typically occur for weakly-interacting exciton-polaritons at high densities. 
Indeed, in this regime polariton nonlinearity shows up with the blueshift~\cite{yagafarov2020mechanisms} that in turn creates a local potential leading to the outflow of particles from the condensate area and broadens the distribution in $k$-space~\cite{PhysRevB.110.045125}. As a result, particles redistribute towards higher-energy states, leading to an increase of the effective temperature in the low-energy fraction (Fig.~\ref{fig:3}b).









\section*{Stimulated cooling}

\begin{figure}
\includegraphics[width=0.95\linewidth]{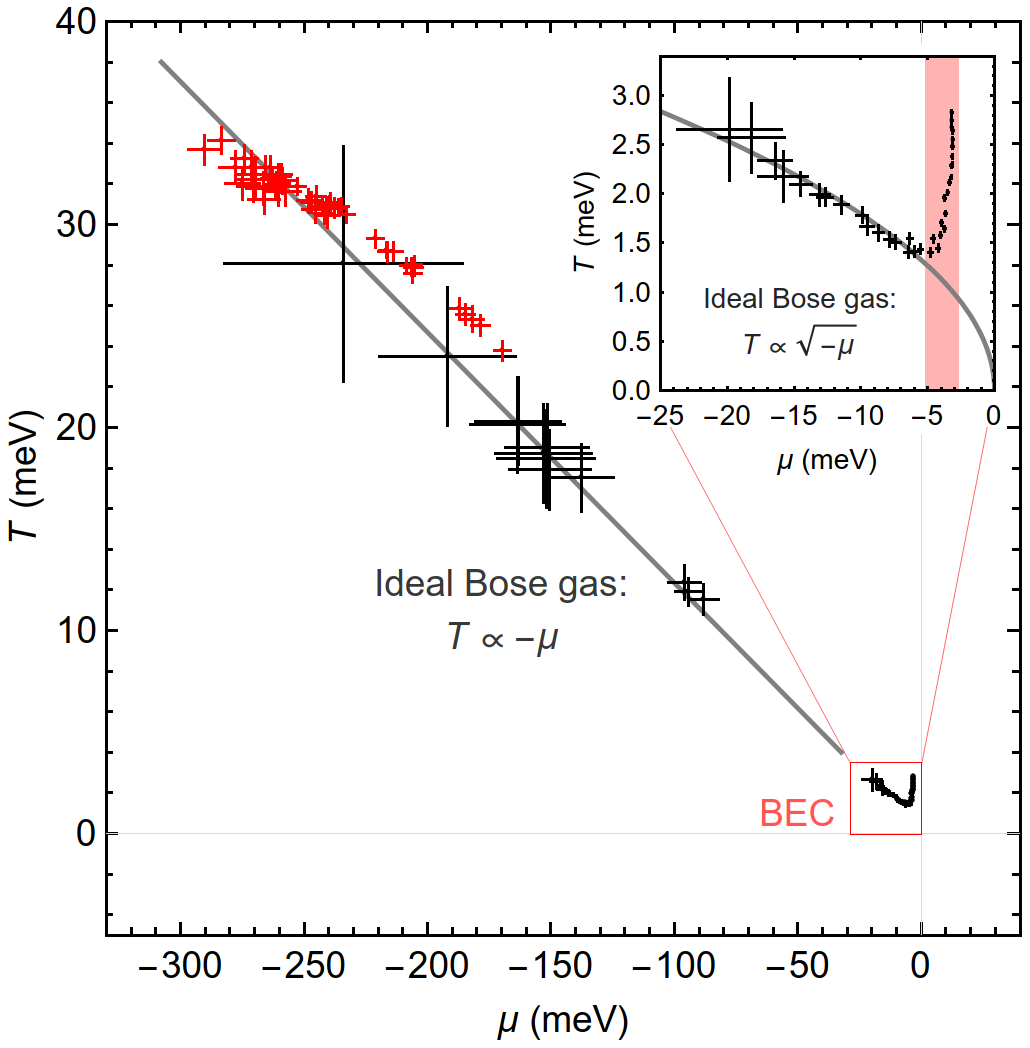}
\caption{
{\bf Universal scaling of thermalization $\bf{T(\mu)}$. }
The temperature of particles ($T$) versus their chemical potential ($\mu$) both obtained from Bose--Einstein fit for the frequency range $(2.55~{\rm eV},~2.57~{\rm eV})$ (black dots) and $(2.585~{\rm eV},~2.61~{\rm eV})$ (red dots).
Horizontal and vertical error bars indicate the standard errors of the fitted temperature and chemical potential, respectively. Gray lines show the $T(\mu)$ dependence predicted for an ideal Bose gas in the  {\it canonical ensemble}.
Inset: Temperature versus chemical potential above the BEC threshold. The red shaded region highlights the part of the $T(\mu)$ dependence not captured by the equilibrium ideal-gas theory, beyond non-interacting particles picture.
Note, the chemical potential of the lower energy fraction at the threshold is $\mu_{th}= -80~{\rm meV}$, according to the date in Fig.~\ref{fig:3}c.
}
    \label{fig:4}
\end{figure}

A similar trend in negative change of the effective temperature has been observed experimentally across light-matter systems~\cite{alnatah2024bose, satapathy2022thermalization}. However, in those experiments the effect stayed relatively close to the bath temperature (typically $\sim 200~{\rm K}$), and its physical origin has remained unclear. Here we observe deep cooling: the polariton gas cools from near room temperature down to sub-bath temperatures below $20~{\rm K}$ (Fig.~\ref{fig:3}b).
An open quantum perspective on this effect shows the cooling effect is the distinct feature of the non-equilibrium nature of the BECs~\cite{PhysRevLett.128.065301}.
To examine stimulated origin directly, we analyze how the effective temperature scales with particle density for both the low-energy and the high-energy fractions. We use the chemical potential as parameter defining the particle density and note that the BEC transition occurs around $\mu_{\rm th} \approx -80~\mu{\rm eV}$, as indicated in Fig.~\ref{fig:3}c for the low-energy fraction. Strikingly, both fractions follow the same dependence $T(\mu)$ over the entire range we access (Fig.~\ref{fig:4}), from well below threshold to several times above it. We observe a smooth transition from low particle density region at high-energy fraction (red dots) to the dense polariton gas at the low energies (black dots) suggesting a universal bosonic mechanism setting the temperature across energy states and densities of the Bose gas. Once the system becomes sufficiently occupied, stimulated processes dominate the redistribution and drive the system into a deeply cooled state.
Importantly, and somewhat unexpectedly, the observed universal (Fig.~\ref{fig:4}) matches the equilibrium statistical theory of an ideal Bose gas in the \textit{canonical ensemble}, namely $\mu \propto -T$ below threshold~\cite{kubo1988statistical}, and $\mu \propto -T^{2}$ above threshold (see Supplemental Information). The temperature vs chemical potential dependence not only supports our suggestion about stimulated origin of the observed cooling, but establishes the connection between non-equilibrium and equilibrium BEC and narrowing the long-standing gap in understanding of how thermodynamic features of light-matter BEC emerge from driven–dissipative dynamics.

When the chemical potential approaches zero (roughly $\mu \gtrsim -5~\mu{\rm eV}$, above $\sim 2P_{\rm th}$), the polariton gas becomes too dense and nonlinear effects start to dominate, drawing the limit for the stimulated cooling of the ideal Bose gas. In this regime the system significantly deviates from the non-interacting ideal-gas picture. Interactions and density-dependent blueshifts reshape potential landscape, enhance outflow from the condensate region, and ultimately lead to heating of the gas, as shown in Fig.~\ref{fig:4} inset (red area). The heating effect cannot be explained within the quantum theory nonequilibrium ideal Bose gas~\cite{PhysRevLett.128.065301}.
Another peculiar aspect in the formation of BEC is the correlation between different polariton states. At the condensation threshold, the second-order correlation is significantly less than unity (see Supplementary Materials). This means that the ground state and the excited states become correlated in such a way that occupations of ground and excited states at the same time are incompatible events. The competition between the ground and nearby excited states accompanied by stimulated transitions gives rise for the correlations between those states.


\section*{Summary}

We reveal stimulated cooling in a driven–dissipative polariton BEC. With increasing particle density, the gas fragments into two thermal fractions occupying different energy ranges: a low-energy fraction that hosts the condensate above threshold, and a high-energy fraction. Both fractions remain well described by Bose–Einstein distributions with distinct effective temperatures and chemical potentials, yet they follow the same universal density scaling: the chemical potential increases while the temperature decreases with particle density. The temperature of the polariton gas, thus, is defined by the chemical potential as $T \propto -\mu $ and $T \propto (-\mu)^{1/2}$, below and above the BEC threshold, respectively. As a result, the low-energy fraction cools deeply, reaching temperatures of order $\sim 20$ K, far below the bath temperature. At the highest densities this behaviour breaks down, as the nonlinear interactions enter the dynamics resulting in outflow of the particles from the condensate region which leads to partial reheating and $k$-space broadening.

\section*{Acknowledgments}

This work was supported by the Finnish Research Impact Foundation within the Tandem Industry Academia (TIA) Seed Project no. 667, by Jane and Aatos Erkko Foundation and the Technology Industries of Finland Centennial Foundation as part of the Future Makers funding program, by the Research Council of Finland under project number 339313, and by the Research Council of Finland through the Finnish Quantum Flagship project 358877. The work is part of the Research Council of Finland Flagship Programme, Photonics Research and Innovation (PREIN), decision number 346529, Aalto University. This work is part of the Finnish Centre of Excellence in Quantum Materials (QMAT). Sh.V.Yu. thanks the Magnus Ehrnrooth foundation. We acknowledge support by the Aalto Science Institute (AScI) International Summer Research Programme.

\subsection*{Methods}

{\bf Sample Fabrication.}
The sample is composed of a Methyl-substituted ladder-type poly(p-phenylene) (MeLPPP; Mn = 31,500) organic material sandwiched between a distributed Bragg reflector cavity. Bottom distributed Bragg reflector (DBR) on a fused silica substrate (mirror 2), a central cavity defect region with an effective thickness slightly larger than half the exciton transition wavelength, and a top DBR (mirror 1). The DBRs consist of alternating SiO2/Ta2O5 quarter-wavelength-thick layers produced by sputter deposition (9+0.5 pairs for the bottom DBR, 6+0.5 for the top DBR). The center of the cavity consists of the polymer layer sandwiched between 50 nm spacer layers of sputtered SiO2. The SiO2 spacer is sputtered on the organic using a SiO2 sputter target. Methyl-substituted ladder-type poly(p-phenylene) (MeLPPP; Mn=31,500 (number averaged molecular weight), Mw=79,000 (weight averaged molecular weight)) was synthesized as described elsewhere. MeLPPP is dissolved in toluene and spin-coated on the bottom spacer layer. The film thickness of approximately 35 nm is measured with a profilometer (Veeco Dektak).

{\bf Experimental Methods}
The sample is excited at 2.76 eV and 45$^\circ$ angle of incidence. The excitation beam generated by tunable optical parametric amplifier (OPA, Light Conversion Topas Prime) with a pulse length of 50 fs and a repetition frequency of 1 kHz. The excitation is tuned in resonance with blue-detuned optomechanical condition in the system \cite{zasedatelev2021single}, with a 50 meV full width at half maximum (FWHM). This OPA is pumped by a 30 fs Ti:Sapphire amplifier (Coherent Astrella). Upon exiting the OPA, we conduct polarization and spatial filtering of the beam via polarizing beam splitters and optical telescopes, before being focused down to a 25 $\mu$m excitation spot.

The dependencies for the BEC are obtained by collecting emission from the sample by a 0.6NA numerical aperture microscope objective (50X Nikon CFI60 TU Plan Epi ELWD Infinity Corrected Objective). A reconfigurable 4f imaging system is constructed, which allows us to observe both the real- and the k-space distribution of the BEC. The emission was focused on the 20 $\mu$m slit of the spectrometer (Oxford Instruments Andor Shamrock 500i), which is equipped with 1200 lines/mm grating and imaged on an EMCCD camera (Oxford Instruments Andor Newton 971). To enhance the imaging and improve signal-to-noise ratio (SNR), the EMCCD gain is adjusted to be 150$\times$, with an exposure time between 1 to 10 seconds depending on the emission strength. 

Single shot measurements are performed using 1 kHz pulse train from Coherent SDG Elite as the `clock'. The clock pulse is used to trigger a FPGA multi-instrument device (Liquid Instruments Moku:Pro), in which the oscilloscope records the pulse intensity from a photodiode 
and a waveform generator sends out a cascaded pulse to the camera to begin acquisition. The exposure time is limited to 0.75 ms and is offset to begin before the excitation pulse. As a result, exposure is limited to capture the single BEC realization generated by a single excitation pulse, with the photodiode recording the pump pulse energy for the respective BEC realization. 


\bibliography{references}

\end{document}